\documentclass[twocolumn]{aastex631}
\usepackage{graphicx}
\usepackage{CJK}
\usepackage{booktabs}
\usepackage{threeparttable}
\usepackage{natbib}

\newcommand{\pdis}{d_{\text{p}}}
\newcommand{\dv}{\Delta v}
\newcommand{\logm}{\log M_\star}

\newcommand{\dsfr}{\Delta \log \text{SFR}}
\newcommand{\logsfr}{\log \text{SFR}}

\newcommand{\kpc}{\ \text{kpc}}
\newcommand{\mpc}{\ \text{Mpc}}
\newcommand{\kms}{\ \text{km}\ \text{s}^{-1}}

\shorttitle{Suppression of Star Formation in Galaxy Pairs}
\shortauthors{Feng et al.}

\begin{document}
\begin{CJK*}{UTF8}{gbsn}

\title{Suppression of Star Formation in Galaxy Pairs}

\correspondingauthor{Shuai Feng, Shi-Yin Shen}
\email{sfeng@hebtu.edu.cn, ssy@shao.ac.cn}

\author[0000-0002-9767-9237]{Shuai Feng (冯帅)}
\affiliation{College of Physics, Hebei Normal University, 20 South Erhuan Road, Shijiazhuang 050024, China}
\affiliation{Guoshoujing Institute of Astronomy, Hebei Normal University, 20 South Erhuan Road, Shijiazhuang 050024, China}
\affiliation{Hebei Key Laboratory of Photophysics Research and Application, Shijiazhuang 050024, China}

\author[0000-0002-3073-5871]{Shi-Yin Shen (沈世银)}
\affiliation{Key Laboratory for Research in Galaxies and Cosmology, Shanghai Astronomical Observatory, Chinese Academy of Sciences, \\ 80 Nandan Road, Shanghai 200030, China}
\affiliation{Key Lab for Astrophysics, Shanghai 200234, China}

\author[0000-0001-6763-5869]{Fang-Ting Yuan (袁方婷)}
\affiliation{Key Laboratory for Research in Galaxies and Cosmology, Shanghai Astronomical Observatory, Chinese Academy of Sciences, \\ 80 Nandan Road, Shanghai 200030, China}

\author[0000-0002-4299-095X]{Wen-Xin Zhong (钟文心)}
\affiliation{Sanda University, No. 2727 Jinhai Road, Shanghai, 201209，China PR}

\author[0000-0003-1359-9908]{Wen-Yuan Cui (崔文元)}
\affiliation{College of Physics, Hebei Normal University, 20 South Erhuan Road, Shijiazhuang 050024, China}
\affiliation{Guoshoujing Institute of Astronomy, Hebei Normal University, 20 South Erhuan Road, Shijiazhuang 050024, China}

\author[0000-0003-1454-2268]{Lin-Lin Li (李林林)}
\affiliation{College of Physics, Hebei Normal University, 20 South Erhuan Road, Shijiazhuang 050024, China}
\affiliation{Guoshoujing Institute of Astronomy, Hebei Normal University, 20 South Erhuan Road, Shijiazhuang 050024, China}

\begin{abstract}
We investigate the suppression of star formation in galaxy pairs based on the isolated galaxy pair sample derived from the SDSS survey. By comparing the star formation rate between late-type galaxies in galaxy pairs and those in the isolated environment, we detect the signal of star formation suppression in galaxy pairs at $\pdis < 100\kpc$ and $200\kpc < \pdis < 350\kpc$. The occurrence of star formation suppression in these late-type galaxies requires their companion galaxies to have an early-type morphology ($n_s > 2.5$). Star formation suppression in wide galaxy pairs with $200\kpc < \pdis < 350\kpc$ mainly occurs in massive late-type galaxies, while in close galaxy pairs with $\pdis < 100\kpc$, it only appears in late-type galaxies with a massive companion ( $\logm > 11.0$), nearly independent of their own stellar mass. Based on these findings, we infer that star formation suppression in wide galaxy pairs is actually a result of galaxy conformity, while in close galaxy pairs, it stems from the influence of hot circum-galactic medium surrounding companion galaxies.
\end{abstract}


\section{Introduction} \label{sec:intro}

In the framework of hierarchical galaxy formation, galaxy merging plays a key role in the mass assembly and structure formation of galaxies. Before the final coalesce, two galaxies are gravitationally bound in the form of galaxy pairs for $1-2$Gyrs \citep{Kitzbichler2008, Lotz2008}. In galaxy pairs, the evolution of member galaxies is inevitably influenced by their companions, resulting in a variety of peculiar properties compared to the galaxies in isolated environments. Nevertheless, how galaxy interaction affects the physical properties of galaxy pairs, especially the star formation rate (SFR), is still not clear. 

The enhancement of star formation is one of the most prominent features of galaxy pairs, which has been known since the 1970s \citep{Larson1978}. Numerous studies have shown that the SFRs of galaxy pairs are significantly higher than those of isolated galaxies at given stellar masses \citep{Barton2000, Ellison2008}. Such an enhancement of star formation shows a strong correlation with the projected separation between pair members, where the enhancement is detectable up to $\pdis \sim 150\kpc$ and reaches the maximum at $\pdis < 30\kpc$ \citep{Li2008, Patton2013, Feng2019}. Furthermore, enhanced star formation is usually correlated with other peculiarities, such as asymmetric photometric morphology \citep{Patton2016, Pan2019}, disturbed gas kinematics \citep{Bloom2018, Feng2020}, and diluted gas phase metallicity \citep{Kewley2006, Scudder2012}. In combination with the result of the numerical simulation, the enhanced star formation of galaxy pairs can be explained by the scenario of tidal-induced gas inflow, where the tidal perturbation from companion galaxies is the key to the nature of star formation \citep{DiMatteo2008, Torrey2012}. 

However, more detailed studies later pointed out that not all galaxy pairs exhibit enhanced star formation, which depends on the types of companion galaxies. They found that galaxies with late-type (star-forming) companions usually display increased SFRs, while it is not for galaxies with early-type (passive) companions for given stellar mass intervals \citep{Park2009, Xu2010, Yuan2012}. This indicates that there may be other physical mechanisms regulating the star formation of paired galaxies. 

In terms of the star formation properties of paired galaxies with early-type (passive) companions, previous studies have not reached an agreement. Some studies suggested that the star formation of these galaxies is comparable to that of isolated galaxies \citep{Cao2016, He2022}, while some other studies considered that their star formation is suppressed compared to that of isolated galaxies \citep{Moon2019, Brown2023}. The divergence in observations led to controversy about the physical mechanisms that regulate the star formation of galaxy pairs, including whether the effects of the hot gaseous halo are not negligible \citep{Hwang2015, Zuo2018, Lisenfeld2019, Moon2019}. 

In addition to the properties of companion galaxies, some other factors, such as the projected separation and mass ratio, are also related to the star formation suppression in galaxy pairs \citep{Davies2015, Ellison2022, Das2023}. Behind these dependencies lie associated physical mechanisms, which are crucial for understanding the star formation properties during galaxy mergers. Clarifying whether star formation suppression exists in galaxy pairs and its dependency relationships are of great significance for further understanding the process of galaxy merging.

In this paper, we investigate the star formation of galaxy pairs based on a large galaxy pair sample, in particular focusing on the suppression of star formation. By analyzing the dependence of the occurrence of star formation suppression, we are attempting to gain insight into the physical process that occurs in paired galaxies. This paper is constructed as follows. We first introduce the galaxy pair sample (Section \ref{sec:sample}), then analyze the dependence of star formation suppression on the properties of galaxy pairs (Section \ref{sec:sfr}). In Section \ref{sec:dis}, we discuss the physical origin of star formation suppression. Finally, we give a summary in Section \ref{sec:sum}. Throughout this paper, we use the standard cosmological model with $\text{H}_0 = 70 \text{km}\ \text{s}^{-1} \text{Mpc}^{-1}$, $\Omega_{\Lambda} = 0.7$, $\Omega_m = 0.3$.

\section{Data}\label{sec:sample}

The galaxy pair sample is obtained from the SDSS DR7 main galaxy sample \citep{SDSSDR7, Strauss2002}, where all galaxies are brighter than $m_r=17.77$. We supplement a significant number of spectral redshifts from other spectral surveys, such as LAMOST \citep{Luo2015} and GAMA \citep{Baldry2018}, to minimize the spectral incompleteness induced by the fiber collision effect (see more details in \citealt{Shen2016} and \citealt{Feng2019}). After supplementation, $713, 366$ galaxies have spectral redshifts and are named the spectroscopic sample). 

We select galaxy pairs from the spectroscopic sample, ensuring that two member galaxies in each pair satisfy the following criteria: (1) the difference of Petrosian magnitude in SDSS-$r$ band is $|\Delta m_r| < 2.5$; (2) the projected separation between them fulfills $10 \kpc < \pdis < 500\kpc$; (3) the difference in line-of-sight velocity satisfies $|\dv| <1000\kms$. 

To ensure that our galaxy pairs are isolated pairs rather than part of galaxy groups/clusters, We require that each member galaxy of a galaxy pair has no other neighbor galaxies except for the other member galaxy. In practice, for a pair member with an apparent magnitude of $m_r$, there should be no other galaxy brighter than $m_r+2.5$ within a range of $\pdis < 500\kpc$ and $|\dv| <1000\kms$ in the spectroscopic sample. Additionally, we also require that there are no other neighbor galaxies around each pair member in the photometric sample in which galaxies have no spectral redshift measurement. The photometric sample contains two parts: galaxies in the SDSS main galaxy sample that is missed by the spectral observations, and galaxies with $17.77<m_r<20.5$. For a pair member with a redshift of $z$ and magnitude of $m_r$, if it has at least one neighbor galaxy brighter than $m_r+2.5$ within a range of $\pdis < 500\kpc$ and $| z-z_{\text{phot}}| < \delta z_{\text{phot}}$ in the photometric sample, then we consider that this galaxy pair is not isolated. The $z_{\text{phot}}$ and $\delta z_{\text{phot}}$ are the photometric redshift and its uncertainty of galaxies in the photometric sample, which are taken from \citet{Beck2016}. In this step, we obtain $13,321$ isolated galaxy pairs. 

For each pair member, we cross-match the stellar mass and global SFR with the MPA-JHU catalog \footnote{\url{https://www.sdss.org/dr17/spectro/galaxy_mpajhu/}} \citep{Brinchmann2004, Kauffmann2003}, and adopt the Sersic index from \citet{Simard2011} to represent galaxy morphology. Finally, we have $11,265$ galaxy pairs with the measurements of stellar mass ($\logm$), star formation rate ($\logsfr$), and Sersic index ($n_s$) for both two pair members. According to the morphology of pair members, we separate galaxy pairs into three categories: (1) $3,275$ S-S pairs, comprising two late-type galaxies (LTG, $n_s<2.5$); (2) $4,918$ S-E pairs, comprising one early-type galaxy (ETG, $n_s \ge 2.5$) and one late-type galaxy; (3) $3,072$ E-E pairs, comprising two early-type galaxies. 

We also identify isolated galaxies from the spectroscopic sample for our following analysis. We define a galaxy with magnitude $m_r$ and redshift $z$ as an isolated galaxy if it satisfies the following two criteria: (1) in the spectroscopic sample, there are no neighbor galaxies brighter than $m_r+2$ located within a range of $\pdis < 500 \kpc$ and $\dv < 1000 \kms$ from it. (2) in the photometric sample, there are no galaxies brighter than $m_r+2$ located within a range of $\pdis < 500 \kpc$ and $|z-z_{\text{phot}}| < \delta z_{\text{phot}}$ from it. With these two criteria, we obtain $12345$ isolated galaxies. 

\section{Result}\label{sec:sfr}

\begin{figure*}
    \centering
    \includegraphics[width=\textwidth]{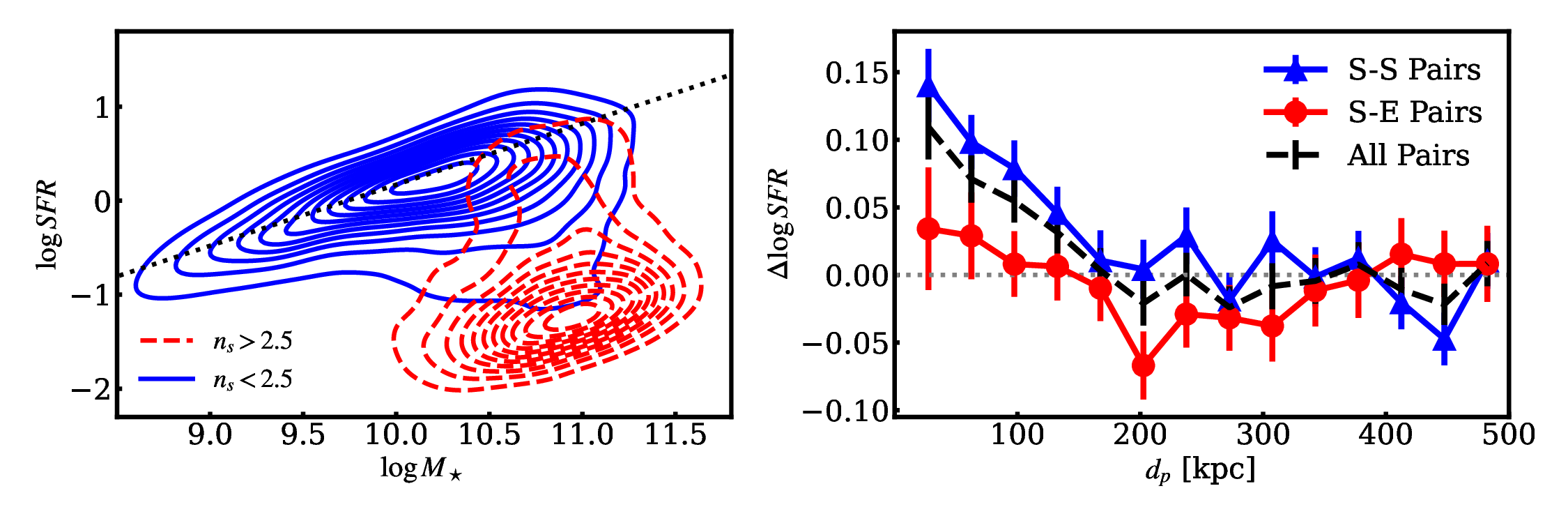}
    \caption{\textit{Left:} SFR as a function of stellar mass for LTGs ($n_s<2.5$, blue contours) and ETGs ($n_s \ge 2.5$, red contours). The black dotted line represents the star-forming main sequence of \citet{Salim2007}. \textit{Right:} The variation of SFR between LTGs in galaxy pairs and isolated galaxies as a function of projected separation for S+S pairs (blue triangles), S+E pairs (red circles), and all pairs (gray solid line). }
    \label{fig:dSFR_Morph}
\end{figure*}

In this study, we do not simply use the star-forming galaxies as the study sample because that may bring about selection effects when discussing the star-forming properties. Instead, we examine the global SFRs of LTGs, which by definition ($n<2.5)$ are not directly related to their star formation properties. As shown in the left panel of Figure \ref{fig:dSFR_Morph}, most of the LTGs (blue solid contours) are star-forming galaxies located in the star-forming main sequence (black dotted lines, adopted from the result of \citealt{Salim2007}). In contrast, ETGs (red dashed contours) fall below the star-forming main sequence. 

In the following, we only consider the S-S pairs and the S-E pairs. For convenience, we use `target galaxy' to denote the pair member whose SFR we aim to study. The other pair member is denoted as a `companion galaxy'. In particular, for S-E pairs, the target galaxy is the late-type pair member, and the companion galaxy is the early-type pair member. In S-S pairs, two late-type pair members take turns to be the target galaxy and companion galaxy once each.

For each LTG in galaxy pairs, we match $5$ isolated galaxies as control galaxies \footnote{Only $32$ LTGs having less than $5$ control galaxies, which are excluded in the following analysis.}. The control galaxies should have the similar stellar mass ($|\Delta \logm|<0.2$), redshift ($|\Delta z| <0.02$) and Sersic index ($|\Delta n_s|<0.2$) as the corresponding paired galaxies. Moreover, we require that control galaxies have morphologies similar to paired galaxies because the SFR is strongly correlated with morphology as shown in the left panel of Figure \ref{fig:dSFR_Morph}. This requirement would prevent us from underestimating the suppression of star formation by preventing false matching of isolated ETGs to the LTGs in galaxy pairs.

We take the median SFR of the control galaxies to represent the characteristic SFR of the isolated galaxies ($\text{SFR}_0$). Then, we define  the variation of SFR (either enhancement or suppression)
\begin{equation}
    \Delta \log \text{SFR} = \log \text{SFR}_{\text{pair}} - \log \text{SFR}_0
\end{equation}
where $\log \text{SFR}_{\text{pair}}$ is the star formation rate of the paired galaxies. In the following sections, we use $\dsfr$ to explore whether and where star formation of paired galaxies is suppressed.

\subsection{Projected Separation}

The right panel of Figure \ref{fig:dSFR_Morph} displays the variation of SFR as a function of the projected separation of the paired members. The gray dashed line shows the median value of $\dsfr$ at the given $\pdis$ intervals for all galaxy pairs. The error bars represent the uncertainty of those median values estimated by the bootstrap method. The general trend of $\dsfr$ for all galaxy pairs (black dashed lines) is nearly the same as in previous studies \citep[e.g.][]{Li2008, Patton2013, Feng2019}. At $\pdis < 150\kpc$, the SFR increases significantly as the projected separation decreases, indicating the influence of galaxy interaction. While at $\pdis > 150\kpc$, the SFR of galaxy pairs is almost identical to that of isolated galaxies. 

However, when we investigate the behavior of $\dsfr$ separately for the S-S pairs (blue triangles) and S-E pairs (red circles), we find that the projected separation, below which galaxy pairs exhibit different star formation properties than isolated galaxies, has reached $350\kpc$. This value is more than twice as large as previous results \citep[e.g. $\sim 150\kpc$ in][]{Patton2016, Feng2019, Feng2020}. Furthermore, we notice that galaxy interaction not only triggers the enhancement of star formation but can also suppress star formation. 

When companion galaxies are ETGs (in S-E pairs), $\dsfr$ of target galaxies have negative values at $\pdis < 350\kpc$, indicating suppressed star formation. The suppression starts from $\pdis \sim 350 \kpc$ and becomes stronger as $\pdis$ decreases. The strongest suppression of star formation occurs at $\pdis \sim 200\kpc$. At $\pdis<200\kpc$, the $\dsfr$ begins to increase as $\pdis$ decreases. For very close galaxy pairs with $\pdis < 30\kpc$, the SFR of S-E pairs is nearly the same as that of isolated galaxies. In comparison, the star formation of LTGs with late-type companions (in S-S pairs) is dominated by the enhancement at $\pdis <150\kpc$. At $\pdis > 150\kpc$, the SFR of S-S pairs are nearly comparable to isolated galaxies. There is no evidence that star formation has been suppressed in the $\dsfr$-$\pdis$ relation of S-S pairs, except for one data point at $\pdis \sim 450\kpc$ which may be attributed to the peculiarities of a few individual galaxies. 

\subsection{Stellar Mass of Pair Members}\label{sec:res_mass}

\begin{figure*}
    \centering
    \includegraphics[width=\textwidth]{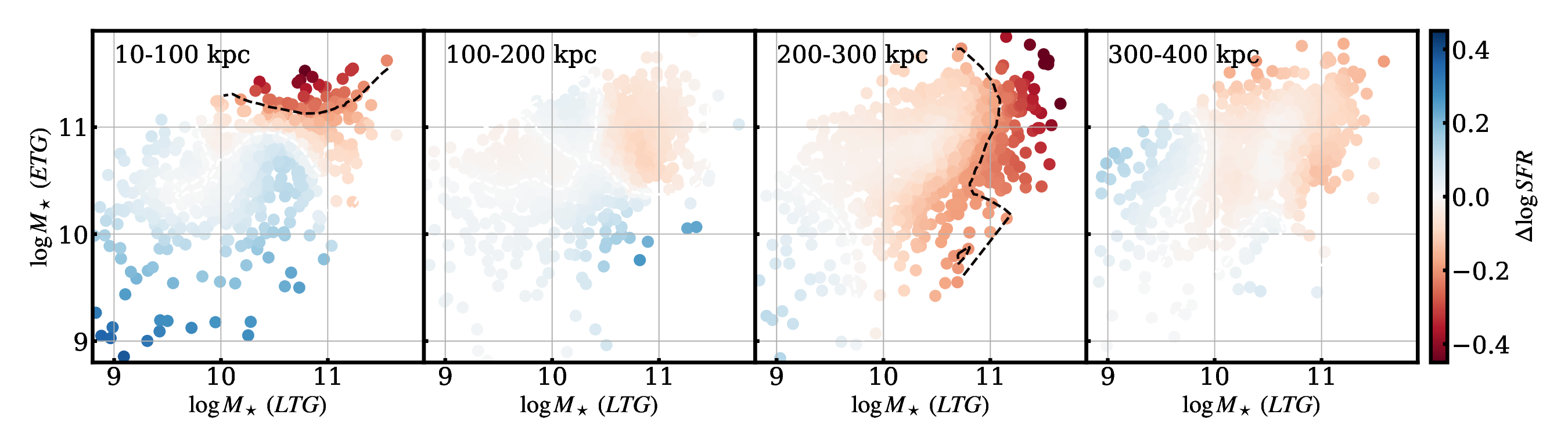}
    \includegraphics[width=\textwidth]{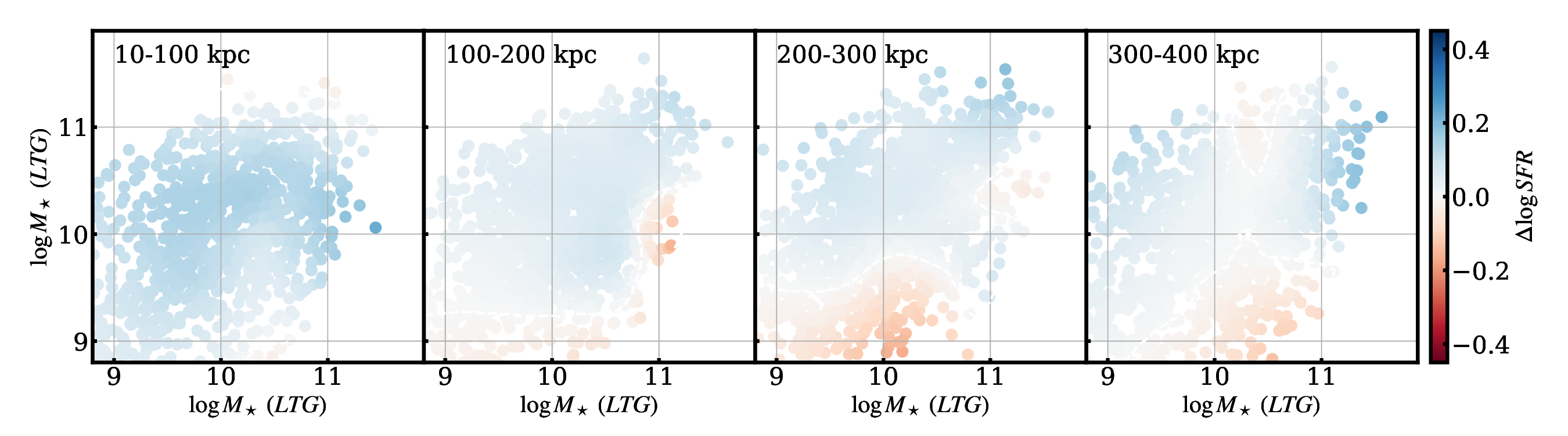}
    \caption{The variation of SFRs as a function of the stellar mass of galaxy pair members. The x-axis shows the stellar mass of the target galaxies, and the y-axis displays the companion galaxies. The color coding represents the variation of SFRs for target galaxies and the black dashed contours represent the $\dsfr = -0.2$. The top rows show the result when companion galaxies are early-type galaxies, whereas the bottom rows show the result when companion galaxies are late-type galaxies. The projected separation range is labeled in the upper left corner of each panel. }
    \label{fig:dSFR_Mass}
\end{figure*}

Many studies demonstrated that the stellar mass of galaxy pairs has a strong impact on the enhancement of star formation \citep[e.g.][]{Scudder2012}, which may also affect the suppression of star formation. To better understand the behavior of star formation suppression, we investigate the relationship between the suppression and the stellar mass of pair members. Particularly, we focus on the galaxy pairs with $\pdis < 400\kpc$, at which the significant suppression of star formation mainly occurs according to the right panel of Figure \ref{fig:dSFR_Morph}.

Figure \ref{fig:dSFR_Mass} displays the $\dsfr$ as a function of the stellar mass of the pair members. The upper panels show the result of S-E pairs, and the lower panels show the result of S-S pairs. Each column represents the different projected separation range, which is labeled in the upper left corner of each panel. In each panel, the x-axis represents the stellar mass of the target galaxies, and the y-axis represents the stellar mass of the companion galaxies. The $\dsfr$ of target galaxies are displayed by color-coded circles. The red color represents the suppression of star formation ($\dsfr < 0$), while the blue color represents the enhancement of star formation ($\dsfr > 0$). To display the general trend, the values of $\dsfr$ are smoothed by the LOESS algorithm \citep{Cappellari2013}. For convenience, we use black dashed lines to label $\dsfr = -0.2$ to better illustrate the dependency of the star formation suppression on the stellar mass. 

These figures reinforce the crucial role of the morphology of companion galaxies and projected separation revealed by Figure \ref{fig:dSFR_Morph}. Firstly, the early-type morphology is more conducive to inducing the suppression of star formation. We find that the significant signals of star formation suppression ($\dsfr < -0.2$) only appeared within the S-E pairs. When comparing the top panels to the bottom panels, it is shown that the S-E pairs are more likely to display negative $\dsfr$ than the S-S pairs with given stellar mass and projected separation. Second, the suppression of star formation is related to the projected separation. At  $200 < \pdis < 300 \kpc$, the S-E galaxy pairs exhibit a remarkably significant suppression of star formation, as demonstrated in Figure \ref{fig:dSFR_Morph}. Additionally, we observed that close galaxy pairs with $\pdis < 100 \kpc$ can also display strong suppression of star formation. This phenomenon is only evident when we analyze galaxy pairs based on stellar mass separately. Hence, the stellar mass of member galaxies is another key factor in the star formation suppression observed in galaxy pairs.

The dependence of the star formation suppression on the stellar mass of member galaxies varies at different projected separations. For close S-E galaxy pairs with $\pdis < 100\kpc$, the suppression of star formation primarily depends on the stellar mass of the companion galaxy. There is a characteristic stellar mass of companion galaxies, approximately $\logm \sim 11.0$, above which the star formation in the target galaxy shows significant suppression. Below this value, the SFR of the target galaxy is enhanced, consistent with previous findings in studies on galaxy pairs \citep{Ellison2008, Li2008, Patton2016, Feng2019}. For close galaxy pairs, both stellar formation suppression and enhancement may occur. Specifically in our galaxy pair sample, the S-E pairs with $\logm > 11.0$ companions display negative $\dsfr$, while the S-E pairs with $\logm < 11.0$ companions display positive $\dsfr$. Due to the relatively low number of S-E galaxy pairs with massive companion galaxies, the signal of star formation suppression is obscured by the signal of enhancement when calculating the average $\dsfr$ of all S-E galaxy pairs with $\pdis < 100\kpc$. Therefore, we cannot observe the star formation suppression of close galaxy pairs in Figure \ref{fig:dSFR_Morph}. 

At $200\kpc<\pdis<300\kpc$, the star formation suppression is primarily determined by the stellar mass of the target galaxy rather than the companion galaxy. According to the results of the top third panel in Figure \ref{fig:dSFR_Mass}, the significant star formation suppression mainly occurs in S-E pairs where the stellar mass of the target galaxy is greater than $\logm \sim 11.0$. The close galaxy pairs and wide galaxy pairs exhibit completely different dependencies, suggesting that the star formation suppression phenomenon in these two types of galaxy pairs may have different physical origins. We speculate that the star formation suppression in close galaxy pairs may originate from the interaction with the hot gas surrounding the companion galaxy (see more discussion in Section \ref{sec:res_nc} and Section \ref{sec:dis_cgm}), while the suppression phenomenon in wide galaxy pairs may be a manifestation of galaxy conformity (see more discussion in Section \ref{sec:dis_gcon}). 

\subsection{Sersic Index of Companion Galaxies} \label{sec:res_nc}

\begin{figure}
    \centering
    \includegraphics[width=\columnwidth]{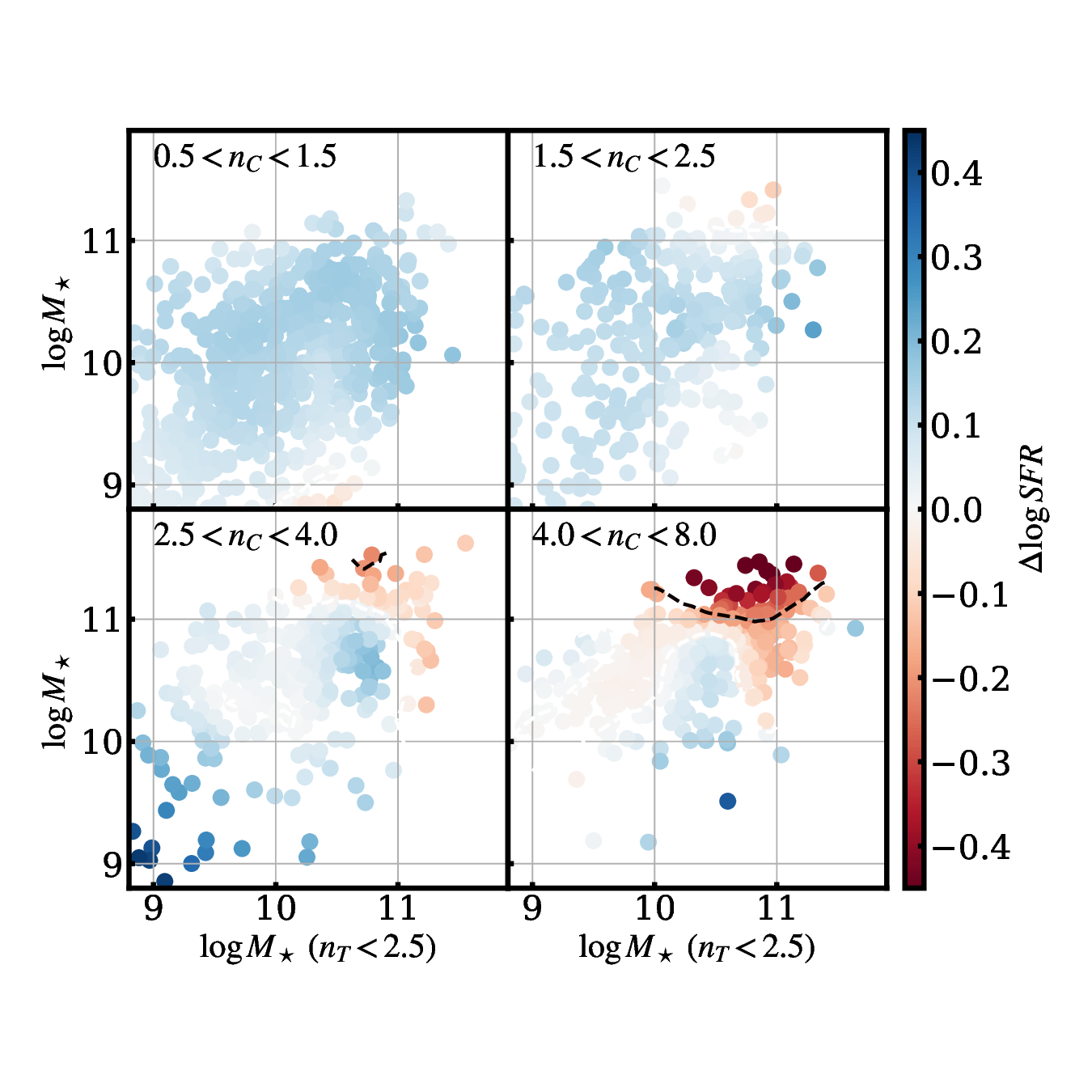}
    \caption{Similar with Figure \ref{fig:dSFR_Mass}. All pairs are $\pdis<100\kpc$, and the target galaxies satisfy $n_T<2.5$. The four panels represent four Sersic index intervals of companion galaxies which are labeled in the top left corner.}
    \label{fig:dSFR_Ns_Comp}
\end{figure}

Compared to LTGs, ETGs are evolved galaxies that have experienced more galaxy merging processes. Because galaxy merging is conducive to the formation of hot gaseous halos, the correlation between the morphology of companion galaxies and star formation suppression at $\pdis < 100\kpc$ may imply that the hot gaseous halos of companion galaxies play an important role in the suppression of star formation. To test this hypothesis, we further examine how the suppression depends on the Sersic index of the companion galaxies which is an indicator of their merging history. 

We divide ETGs and LTGs in companion galaxies into two subcategories according to the Sersic index ($n_C$) respectively and display the results in Figure \ref{fig:dSFR_Ns_Comp}. In this figure, the Sersic index of target galaxies ($n_T$) is smaller than $2.5$ consistent with Figure \ref{fig:dSFR_Mass}, and the range of the Sersic index of companion galaxies is labeled in the upper left corner of each penal. 

These four plots clearly demonstrate the close correlation between the morphology of the companion galaxy and the suppression of star formation in the target galaxy. The higher the Sersic index of the companion galaxy, the stronger the suppression of star formation in the target galaxy. When $n_C$ is less than $1.5$, the star formation in all target galaxies is enhanced. When $1.5 < n_C < 2.5$, the trend of enhanced star formation in the target galaxies, whose companion galaxies satisfy $\logm > 11.0$, is significantly weakened. When $n_C$ is greater than $2.5$, the target galaxies with companion galaxies of stellar mass greater than $\logm \sim 11.0$ begin to exhibit negative values in the $\dsfr$. For the galaxy pairs with the highest $n_C$ values ($n_C > 4.0$), the suppression of star formation is the strongest. Among them, for the target galaxies with companion galaxies having $\logm > 11.0$, the value of $\dsfr$ has already dropped below $-0.2\text{dex}$. Based on this phenomenon, we conclude that the more evolved companion galaxies are more likely to trigger the suppression of star formation in target galaxies. 

We also analyzed the correlation between the Sersic index of the companion galaxies and the suppression of star formation when the Sersic index of the target galaxies is less than $1.5$. The results remained unchanged. Therefore, the correlation observed in this section indeed originates from the morphology of the companion galaxies rather than the morphology of the target galaxies.

\subsection{Morphology of Target Galaxies}

\begin{figure}
    \centering
    \includegraphics[width=\columnwidth]{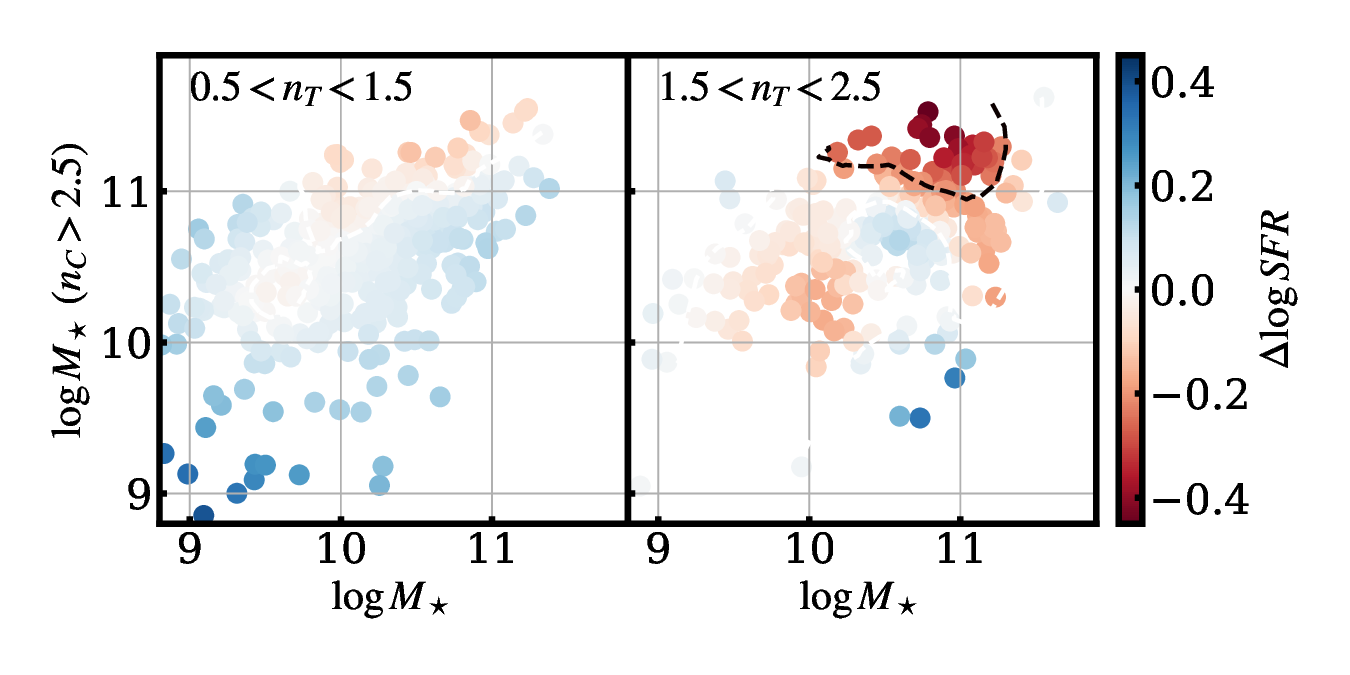}
    \caption{Similar with Figure \ref{fig:dSFR_Mass}. All pairs are $\pdis<100\kpc$, and the companion galaxies satisfy $n_C>2.5$. The two panels represent two Sersic index intervals of target galaxies which are labeled in the top left corner.}
    \label{fig:dSFR_Ns_Tar}
\end{figure}

The morphology of target galaxies can also affect star formation properties of themselves during the merging process according to the previous studies \citep{Barnes1996, DiMatteo2008}. In this section, we investigate the role of the target's morphology in the suppression of star formation. We display the results in Figure \ref{fig:dSFR_Ns_Tar}, where all companion galaxies satisfy $n_C>2.5$ \footnote{Similarly, we also discussed the case when the Sersic index of the companion galaxies is greater than $4.0$, and the final results remained unchanged.}. The range of the Sersic index of target galaxies ($n_T$) is labeled in the upper left corner of each panel. 

Comparing these two panels, we find that the Sersic index of target galaxies indeed influences the properties of star formation. Firstly, for galaxy pairs with companion galaxies $\logm > 11.0$, the larger the Sersic index of the target galaxy, the stronger the suppression of star formation. Target galaxies with $n_T < 1.5$ exhibit only very weak signals of star formation suppression, while those with $n_T > 1.5$ have $\dsfr <-0.2$. Secondly, for galaxy pairs with companion galaxies $\logm < 11.0$, the larger the Sersic index of the target galaxy, the weaker the enhancement of star formation. We suggest that the lower Sersic index of target galaxies is conducive to enhanced star formation, which can weaken the phenomenon of star formation suppression (see more discussion in Section \ref{sec:dis_bulge}).

\section{Discussion}\label{sec:dis}

\subsection{Comparison to Previous Studies}

In general, our results do not conflict with previous studies. Because of the different criteria for selecting galaxy pairs and different methods for analyzing star formation, our study unveils some new phenomena. 

Our galaxy pair sample is selected from the SDSS survey and spans a wide range of stellar mass and mass ratio between pair members, which enables us to explore the dependence on stellar mass in detail. If we focus on the close major merger pairs with $\logm > 10.0$ (similar to the galaxy pairs in \citealt{Xu2010}), the average SFR of the S-E pairs is also comparable to the isolated galaxies, in agreement with previous studies \citep{Xu2010, Yuan2012, Cao2016}. As we discussed in Section \ref{sec:res_mass}, that is actually the combination of the enhancement in target galaxies with $\logm < 11.0$ companions and suppression in target galaxies with $\logm > 11.0$ companions. 

In our study, we analyzed the SFR of LTG rather than star-forming galaxies. This avoids underestimating star formation suppression due to the omission of galaxies whose star formation has already been suppressed. In some previous studies, the suppression of star formation unveiled by the SFR of star-forming galaxies is very limited, but the suppression of star formation can still be reflected by estimating the fraction of quiescent galaxies \citep{Moon2019}. Therefore, these previous results on the suppression of star formation are generally consistent with our study. 

At last, we should mention that most paired galaxies showing suppressed star formation are still star-forming galaxies rather than quiescent galaxies. On one hand, the results of $\dsfr$ in Figure \ref{sec:res_mass} indicate that the degree of star formation suppression does not exceed $0.5\text{dex}$. On the other hand, the distribution of $\text{sSFR}$ for galaxies exhibiting star formation suppression also supports this point. We analyzed the distribution of $\text{sSFR}$ for target galaxies in S-E galaxy pairs with $\pdis < 100 \kpc$, and found that $75\%$ of target galaxies with $\dsfr < -0.2$ still have $\text{sSFR}$ greater than $-11.5$, indicating that they are still star-forming galaxies. 

\subsection{Physical Origin of Star Formation Suppression in Close Galaxy Pairs}

The complicated dependence mentioned in the above sections shows that we cannot explain the star formation suppression in galaxy pairs with a single mechanism. In the following, we discuss two possible physical mechanisms, each of which can explain a part of the observed phenomena, but we do not rule out the contribution of other mechanisms. 

\subsubsection{Circum-galactic Medium around Companion Galaxies}\label{sec:dis_cgm}

The hot circum-galactic medium (CGM) around companion galaxies is one of the explanations for the star formation suppression in target galaxies \citep{Park2009, Moon2019}. Theoretical studies suggest that massive passive galaxies are surrounded by hot gaseous halos, which can shut off the cold gas supply \citep{Dekel2006}. Because the hot halo can extend to the virial radius, it is possible that neighbor galaxies will also be affected by the hot gas. Numerical simulations found that the existence of hot gaseous halos is indeed able to induce the star formation of merging galaxies lower than the isolated galaxies \citep{Monster2011, Karman2015, Hwang2015}. 

In observation, the dependence of hot CGMs on the stellar mass and morphology is highly similar to that of star formation suppression. First, the hot gas content in CGMs is correlated with the morphology of their host galaxies. Both individual detection and statistical studies through the stacking method find that diffuse X-ray emission around ETGs is more luminous than that of LTGs \citep{Anderson2013, Li2017}, indicating that ETGs prefer to hold hotter gaseous halos. In contrast, CGM around LTGs contains more cold gas \citep{Lan2020}. Second, the hot gas content shows a positive correlation with the stellar mass of host galaxies \citep{Anderson2013, Goulding2016, Forbes2017}. There is also a characteristic stellar mass around $\logm \sim 11.0$, below which the signal of hot gas in CGMs is almost undetectable \citep{Greco2015, Comparat2022}. This similarity indicates the suppression of star formation in target galaxies may be induced by the hot gas around companion galaxies. 

Based on the above discussion, we speculate that during close encounters of galaxy pairs, besides the dynamical effects of cold gas, the CGM properties of the companion galaxy can also significantly influence the star formation of galaxies. If the CGM of the companion galaxy contains a large amount of hot gas, the inflow of cold gas in the target galaxy will be greatly weakened, leading to the suppression of star formation.

\subsubsection{Morphology of Target Galaxies} \label{sec:dis_bulge}

The structure of galaxies can influence the properties of star formation during the galaxy merging process. Numerical simulations predicted that massive bulges in disk galaxies can stabilize the gas distribution and suppress the starburst during galaxy encounters by preventing tidal-induced gas inflow \citep{Barnes1996, DiMatteo2008}. Observational studies confirm that the SFR of paired galaxies is indeed strongly correlated with the bulge-to-total ratio, where only galaxies with a low bulge-to-total ratio will exhibit enhancement of star formation \citep{He2022}. In our study, we find that the target galaxy with a lower Sersic index, which approximately equals a smaller bulge-to-total ratio, more easily exhibits enhancement of star formation instead of suppression, consistent with previous findings. 

Combined with the above results, we suggest that the enhancement of star formation and the suppression of star formation is dominated by two independent processes that exist simultaneously. The morphology of target galaxies regulates the enhancement of star formation by stabilizing the existing gas within the disks of target galaxies which can affect the inflow of cold gas, while the morphology and stellar mass of companion galaxies regulate the suppression of star formation by shutting off the cold gas supply through the effect of hot gaseous halos. For $n_s < 1.5$ galaxies with early-type companions, the enhanced star formation caused by their own morphology will offset the suppression of star formation caused by their companion galaxies, resulting in reduced suppression of star formation in those galaxies.

\subsection{Galactic Conformity} \label{sec:dis_gcon}

Galactic conformity, first proposed by \citet{Weinmann2006}, refers to the phenomenon where the star formation properties of satellite galaxies within a galaxy group tend to be similar to those of the central galaxy. If the central galaxy is an early-type galaxy, satellite galaxies often are quiescent as well. This phenomenon is quite similar to the dependency observed in our findings regarding star formation suppression. In many studies on galactic conformity \citep[e.g.][]{Weinmann2006, Phillips2014}, the galaxy group samples typically included a large number of galaxy groups with only two member galaxies. Therefore, galactic conformity also exists in galaxy pairs. Because the galaxy interactions between pair members typically do not extend beyond $\pdis \sim 150 \kpc$ \citep[e.g.][]{Patton2016, Feng2020}, we infer that the observed star formation suppression in wide galaxy pairs with $200\kpc < \pdis < 350\kpc$ may indeed be a manifestation of galactic quenching. 

It is generally believed that galactic conformity arises from the properties of host dark matter halos of galaxy groups, such as the assembly history of dark halos. Therefore, the premise for two galaxies to exhibit galactic conformity is that they are located within the same dark matter halo, in other words, they are gravitationally bound. Based on the relationship between stellar mass and dark halo mass proposed by \citet{Velander2014} and the assumption of a spherical dark matter halo, a galaxy with $\logm \sim 11.0$ has a virial radius of $300\kpc$. Other galaxies within $\pdis < 300 \kpc$ would fall into the host dark halo of this galaxy, thus altering their star formation properties. 

Specifically regarding the results in Figure \ref{fig:dSFR_Mass}, at $\pdis < 300 \kpc$, galaxy pairs containing galaxies with $\logm > 11.0$ are gravitationally bound, and the member galaxies exhibit galactic conformity due to the influence of the properties of the host halo, thereby showing star formation suppression. However, when $\pdis > 300 kpc$, except for a few galaxy pairs with very massive pair members (e.g. $\logm > 11.5$), two member galaxies are not located within the same dark halo, and therefore, star formation suppression does not naturally occur. At $\pdis < 200 \kpc$, the interactions between galaxy pairs can enhance star formation \citep{Patton2016, Feng2020}. As a result, the degree of star formation suppression is weakened at $100\kpc < \pdis < 200 \kpc$.

\section{Summary}\label{sec:sum}

In this paper, we investigate the suppression of star formation in galaxy pairs based on the isolated galaxy pair sample derived from the SDSS main galaxy sample. We select $13,321$ isolated galaxy pairs with the criteria of $|\Delta m_r| < 2.5$, $\pdis<500\mpc$ and $|\dv|<1000\kms$. For a pair member with a magnitude of $m_r$, there are no other neighbor galaxies brighter than $m_r + 2.5$ within $\pdis<500\mpc$ and $|\dv|<1000\kms$. 

By comparing the SFR of late-type pair members ($n_s<2.5$, denoted as target galaxies) to isolated galaxies, we find the following results about the star formation suppression. 

\begin{enumerate}
    \item The suppression of star formation depends on the projected separation between pair members. The significant signal of the suppressed star formation mainly happens in galaxy pairs at $\pdis < 100\kpc$ and $200\kpc < \pdis < 350\kpc$. 

    \item The occurrence of star formation suppression requires the companion galaxy's Sersic index satisfying $n_s > 2.5$, and the degree of suppression increases as the Sersic index of the companion galaxy increases. For target galaxies with a companion galaxy Sersic index less than $2.5$, there is no evidence of star formation suppression.
    
    \item In close galaxy pairs ($\pdis < 100\kpc$), the suppression of star formation further requires the companion galaxy to have a large stellar mass ($\logm > 11.0$). For target galaxies with a companion galaxy $\logm < 11.0$, the star formation is dominated by the enhancement. Besides, the morphology of target galaxies can also influence the suppression of star formation. The star formation of target galaxies with $1.5 < n_s < 2.5$ is more easily suppressed than those with $n_s < 1.5$. 
    
    \item In wide galaxy pairs ($200\kpc < \pdis < 300\kpc$), the suppression of star formation requires the target galaxy to have a large stellar mass ($\logm > 11.0$). 
\end{enumerate}

According to the dependency of star formation suppression, we discuss the possible origin of this phenomenon. We suggest that the suppression of star formation in close galaxy pairs might be the combined effect of the hot circum-galactic medium around companion galaxies and the structure of target galaxies. In wide galaxy pairs, star formation suppression primarily arises from galaxy conformity. To further investigate the physical mechanisms underlying this phenomenon, multi-wavelength studies and numerical simulations are necessary. 

\section*{Acknowlegements}
We thank the anonymous referee for the helpful and constructive comments that improved the paper. This work is supported by the National Natural Science Foundation of China (No. 12103017, 12073059, 12141302, 12173013, 11903012), Natural Science Foundation of Hebei Province (No. A2021205001, A2021205006, A2019205166), Postdoctoral Research Program of Hebei Province (No. B2021003017), the project of Hebei Provincial Department of Science and Technology (No. 226Z7604G) and Science Foundation of Hebei Normal University (No. L2021B08). SF acknowledges the financial support from the Physics Postdoctoral Research Station at Hebei Normal University. SSY acknowledges support from the China Manned Space Project with NO. CMS-CSST-2021-A07, the program of Shanghai Academic/Technology Research Leader (22XD1404200), and the National Key R$\&$D Program of China (No. 2019YFA0405501, 2022YFF0503402). FTY acknowledges support from the Natural Science Foundation of Shanghai (Project Number: 21ZR1474300), the Funds for Key Programs of Shanghai Astronomical Observatory, and the science research grants from the China Manned Space Project with No. CMS-CSST-2021-A04, CMS-CSST-2021-B04. 

\bibliography{ref}
\bibliographystyle{aasjournal}

\end{CJK*}
\end{document}